\documentclass[conference,a4paper]{IEEEtran}
\usepackage[utf8x]{inputenc}
\usepackage[T1]{fontenc}
\usepackage[english]{babel}
\usepackage{color}
\usepackage{verbatim, amsmath, amsfonts, amssymb, amsthm, mathrsfs}
\usepackage{enumerate}
\usepackage{tikz}
\usetikzlibrary{through,intersections}

\usepackage{fixme} %
\usepackage{cite}
\usepackage[kerning=true]{microtype} 
\usepackage[framemethod=TikZ]{mdframed}
\usepackage{bold-extra}

\newtheorem{theorem}{Theorem}
\newtheorem{lemma}{Lemma}
\newtheorem{corollary}{Corollary}
\newtheorem{proposition}{Proposition}

\newtheorem{example}{Example}

\newtheorem{definition}{Definition}
\newtheorem{remark}{Remark}

\newcommand{\ie}{\,\emph{i.e.},\;}

\newcommand*{\lon}{
       \mskip1mu
        \relax
        {:}
        \mskip1mu
        \relax
}

\newcommand{\Ra}{\Rightarrow}

\title{Equivalence of Two Techniques for Proving Non-Shannon-type Inequalities}
\title{Equivalence of Two Proof Techniques for Non-Shannon-type Inequalities}

\author{
  \IEEEauthorblockN{Tarik~Kaced}
  \IEEEauthorblockA{LIRMM (UMR 5506), Université de Montpellier 2, and\\
   Institute of Network Coding, The Chinese University of Hong Kong, Shatin, N.T.\\
    email: tarik@inc.cuhk.edu.hk}}

\colorlet{problem}{green!50!yellow}
\colorlet{todo}{red}
\colorlet{toreview}{blue}
\colorlet{standard}{gray}
\colorlet{remark}{green!60!blue}
\colorlet{tokeep}{black!10!violet}
\colorlet{ok}{black}

\begin{document}
\maketitle

\begin{abstract}

\color{ok}
We compare two different techniques for proving non-Shannon-type information inequalities. 
The first one is the original Zhang-Yeung's method, commonly referred to as the copy/pasting lemma/trick. The copy lemma was used to derive the first conditional and unconditional non-Shannon-type inequalities. The second technique first appeared in Makarychev~\emph{et al} paper \cite{MMRV} and is based on a coding lemma from Ahlswede and Körner works.  
We first emphasize the importance of balanced inequalities and provide a simpler proof of a theorem of Chan's for the case of Shannon-type inequalities. We compare the power of various proof systems based on a single technique.
\end{abstract}


\section{Introduction}

{\color{ok}
Information inequalities are linear inequalities for the Shannon entropy of random variables. They play a central role in information theory for they tell us how much can information be compressed, and are useful in many converse coding theorems.
Determining all the inequalities satisfied by the joint entropy, and thus describing the so-called space of entropic vectors, has become a major challenge in information theory.
Apart from evident applications in all kinds of information-theoretic problems, more fundamental connections are known to exist with matroid theory, Kolmogorov complexity, determinantal inequalities, combinatorics, or group theory. 

Shannon's seminal works \cite{shannon-communication1,shannon-communication2} of the 1940's  introduced, amid many other things, the first information inequality commonly called \emph{the basic inequality}:
\begin{equation*}
H(AC) + H(BC) \ge H(ABC) + H(C). 
\end{equation*}
Which, in the language of information theory, means that the conditional mutual information $I(A\lon B| C)$ is non-negative.

{\color{ok}
Positive linear combinations of instances of the basic inequalities are called \emph{Shannon-type} inequalities. The question of whether these Shannon-type inequalities are {the only valid ones} or not was raised by Pippenger \cite{pippenger} in 1986, yet only answered more than 10 years later. The first non-Shannon-type inequality was proven by Z.~Zhang and R.~W.~Yeung in \cite{ZY98} using the copy trick. Their technique has subsequently been used to find infinite families of non-Shannon-type inequalities (see \cite{matus-inf, SixInequalities, projection-method}).
A few years later, a different technique was discovered by K.~Makarychev, Y.~Makarychev, A.~Romashchenko and N.~Vereshchagin  (see \cite{MMRV}) based on results on sub-achievable entropy vectors for the entropy characterization problem (see \cite[p.~352]{csiszarkorner}). This new technique proved a 5-variable generalization of the original 4-variable Zhang-Yeung inequality. }

To the author's knowledge, these two techniques are the only ones known, to-date, for  proving non-Shannon-type inequalities. The aim of this paper is to study and compare the power of these two techniques. The task of proving Shannon-type inequalities is known to be a LP problem and is better left to computer programs, e.g. ITIP \cite{ITIP} or Xitip \cite{XITIP}. Rules could be added to these programs for the derivation of non-Shannon-type inequalities using the two techniques we mentioned.
 
We show that each   technique can prove the same inequalities modulo rewriting inequalities in some equivalent form. Indeed, a result of Chan's work in \cite{chan-balanced} states that every information inequality can be equivalently put in balanced form. We present an elementary proof of this result for the particular case of Shannon-type inequalities, and argue that balanced inequalities play an important role in the comparison of the two techniques.

{\color{ok}
After fixing notations, the rest of the paper is organized as follows.
Section~\ref{sec:balanced} explains Chan's balanced inequalities, Zhang-Yeung and Makarychev~\emph{et~al} respective techniques are presented in 
Section~\ref{sec:techniques}. Various proof systems involving the two different techniques are compared in Section~\ref{sec:comparison}.
}

\subsection{Preliminaries}

Let $\{X_i\}_{i\in\mathcal{N}}$ be a collection of random variables indexed by a set $\mathcal{N}$ of $n$ elements. For a non-empty subset $J\subseteq\mathcal{N}$, we denote by $X_J$ the set of random variables 
$\{X_j:j\in J\}$.


An unconditional linear information inequality for a set of $n$ random variables
is a linear form with $2^n-1$ real coefficients
$(c_J)_{\varnothing\neq J\subseteq\mathcal{N}}$ such that  for all jointly distributed
random variables $\{X_i\}_{i\in\mathcal{N}}$,
\begin{equation*}
\sum\limits_{\varnothing\neq J\subseteq \mathcal{N} }{c_J H(X_J)} \ge 0.
\end{equation*}

We call \emph{Shannon-type} the inequalities of the set of all positive linear combinations of instances of the basic inequality. That is, a valid inequality that can be put in the form 
\begin{equation}
\sum\limits_{\substack
{\varnothing\neq J\subseteq \mathcal{N}\\
\varnothing\neq K\subseteq \mathcal{N}\\
\varnothing\neq L\subseteq \mathcal{N} }}{c_{J,K,L} I(X_{J}\lon X_{K}| X_{L}) \ge 0},
\label{ineq:shannontype}
\end{equation}
where all $c_{J,K,L}$ are non-negative.

\section{Balanced Inequalities}
\label{sec:balanced}

\begin{definition}[Balanced Inequalities]
An $n$-variable information inequality is said \emph{balanced for variable} $X_i$ if the sum of the coefficients involving $X_i$ is zero:
\begin{equation*}
\sum\limits_{i\in J\subseteq \mathcal{N}}{c_J} = 0.
\end{equation*}
An $n$-variable information is simply called \emph{balanced} if it is balanced for all of its $n$ variables.
\end{definition}

{Given a valid linear information inequality, can one obtain a balanced counterpart that is also a valid information inequality?}  This question was answered in a paper of Chan's (see \cite{chan-balanced}).

\begin{theorem}[Balanced Inequalities, Chan \cite{chan-balanced}]\label{theo:balanced}
Let $(c_J)_{\varnothing\neq J\subseteq \mathcal{N}}$ be a list of coefficients, the following are equivalent:
\begin{enumerate}
\item The inequality
\[
\sum_{\varnothing\neq J\subseteq\mathcal{N}} {c_J H(X_J)}  \ge 0
\]
is a valid information inequality.
\item The inequality
\[
\sum_{\varnothing\neq J\subseteq\mathcal{N}  } {c_J H(X_J)} -
\sum_{ \jmath \in\mathcal{N}  } {r_\jmath H(X_\jmath| X_{\mathcal{N}-\jmath}}) \ge 0,
\]
where $r_\jmath$ is the sum of all $c_J$ involving $\jmath$, 
is a valid balanced information inequality.
\end{enumerate} 
\end{theorem}
The previous result states that any information inequality can be \emph{balanced} by subtracting the corresponding terms. Obviously, the coefficients $r_\jmath$ must be non-negative, hence the balanced inequality appears to be stronger.

{\color{ok}
\begin{example} The $3$-variable inequality
\[ H(X_2,X_3) \ge0 \]
balances into the following inequality
\[ I(X_1\lon X_2 X_3) + I(X_2\lon X_3|X_1) \ge 0. \]
\end{example}
}

The original proof of Theorem~\ref{theo:balanced} involves a random coding argument and Chan-Yeung's technique of entropic vectors approximation using quasi-uniform distributions (see \cite{ChanYeung}). While the original proof is quite involved, we present hereafter a simpler proof for the case of Shannon-type inequalities.

\subsection{Balancing the Basic Inequality}
\label{ssec:balancebasic}


For a set of $n$ random variables $X_\mathcal{N}$, an instance of the basic inequality has the form: 
\begin{equation}
I(X_{J}\lon X_{K}| X_{L}) \ge0,
\label{ineq:basic}
\end{equation}
for $J,K,L$ nonempty subsets of $\mathcal{N}$.

Notice first that inequality \eqref{ineq:basic} is already balanced whenever $J,K,L$ are pairwise disjoint. It is also balanced for any single variable in the set $X_L$ for they appear in each term of the inequality (twice with coefficient $1$ and twice with coefficient $-1$).
For a variable $x$ in $X_J$, the inequality \eqref{ineq:basic} is balanced for $x$ iff $x$ does not appear in $B$. A symmetric remark holds for variables in $X_K$. 
Therefore, the basic inequality \eqref{ineq:basic} is balanced iff $J\cap K = \varnothing$. If $W=J\cap K$ is non-empty, inequality \eqref{ineq:basic} rewrites to:
\begin{align*}
H(X_{W}| X_{L}) + I(X_{J-W}\lon X_{K-W}| X_{W\cup L}) \ge0,
\end{align*}
which balances into
\begin{align*}
I(X_{W}\lon X_{\mathcal{N}-W}| X_{L}) + I(X_{J-W}\lon X_{K-W}| X_{W\cup L}) \ge0.
\end{align*}
This inequality is the sum of two (other) instances of the basic inequality, it is thus a valid Shannon-type inequality. So we have just proven the following proposition:

\begin{proposition}
Theorem~\ref{theo:balanced} holds for instances of the basic inequality.
\label{prop:basic}
\end{proposition}

\begin{example}
For $3$-variables information inequalities, the only balanced instances of the basic inequality are the following ones:
\begin{IEEEeqnarray*}{RCRCRC}
I(X_1\lon X_2X_3) &\ge0, &I(X_2\lon X_1X_3)&\ge0, &I(X_3\lon X_1X_2)&\ge0,\\
I(X_1\lon X_2|X_3) &\ge0, &I(X_1\lon X_3|X_2)&\ge0, &I(X_2\lon X_3|X_1)&\ge0,\\
 I(X_1\lon X_2)&\ge0, &I(X_1\lon X_3)&\ge0, &I(X_2\lon X_3)&\ge0.
\end{IEEEeqnarray*}
(Note that we can recover the first line from the last two.)
\end{example}

\subsection{Balancing Shannon-type Inequalities}

{\color{ok} 
By definition, a Shannon-type inequality (of the form \eqref{ineq:shannontype}) is simply a (weighted) sum of instances of the basic inequality. Since the balanced property is stable by sums, balancing a Shannon-type inequality is the same as balancing each of the instances of the basic inequality in \eqref{ineq:shannontype}. By Proposition~\ref{prop:basic}, the balanced inequality thus obtained is valid:

\begin{corollary}
Theorem~\ref{theo:balanced} holds for Shannon-type information inequalities.
\end{corollary}

Notice that the argument of Subsection~\ref{ssec:balancebasic} shows that the balanced inequality we obtain is Shannon-type. 
}

\subsection{Balancing General Information Inequalities}

{\color{ok} 

For a general (non-Shannon-type) information inequality, we should still rely on the original proof of Theorem~\ref{theo:balanced}, though a more direct proof is not excluded.
Note, however, that most of, if not all, the known non-Shannon-type inequalities are already balanced. 

}

\begin{remark}
Checking if a given inequality is balanced and balancing an inequality have linear complexity in the length of the inputted inequality (as a sum of joint entropies).
\end{remark}

\section{Techniques for non-Shannon-type inequalities}
\label{sec:techniques}

We describe the two main techniques for proving non-Shannon-type information inequalities.

   
\subsection{Zhang-Yeung's Technique}

\begin{mdframed}[%
        frametitle={Rule ZY},
        frametitlefont=\bfseries\scshape,
        skipabove=1pt,
        skipbelow= 3pt,
       innerlinewidth=.5pt,
        frametitlerule=true,
        frametitlebackgroundcolor=gray!30, 
        backgroundcolor=gray!10, 
        nobreak=true]
\begin{enumerate}[(A)]
\item
 If we have an information inequality of the form:
 \begin{equation*}
 f(X_\mathcal{N},Y_\mathcal{M}) + g(Y_\mathcal{M},Z) + 
 \alpha I(Z\lon X_\mathcal{N}|Y_\mathcal{M})\ge 0,
   \end{equation*}
 for some $\alpha\ge0$;
\item then the following (stronger) inequality is also valid:
\[
 f(X_\mathcal{N},Y_\mathcal{M}) + g(Y_\mathcal{M},Z)  \ge 0.
\]
\end{enumerate}
\end{mdframed}
The correctness of this rule is based on the following  lemma.

\begin{lemma}[Copy lemma, \cite{SixInequalities}]\label{lemma:copylemma}
  Let $A,B,C$ be three jointly distributed random variables. There exists a fourth random variable $A'$ such that:
  \begin{itemize}
  \item  $(A,B)$ and $(A',B)$ have the same distribution;
  \item $A'$ is independent of $(A,C)$ given $B$. 
  \end{itemize}Such an $A'$ is called a \emph{$C$-copy of $A$ over $B$}.
\end{lemma}

\begin{IEEEproof}[Proof of Correctness of \textsc{Rule~ZY}]
 Take $Z'$ to be a $X_\mathcal{N}$-copy of $Z$ over $Y_\mathcal{M}$, and apply the inequality of step (A) for $Z=Z'$. By Lemma~\ref{lemma:copylemma}, we obtain the inequality of step (B).
\end{IEEEproof}

This technique has been extensively used to obtain constrained and unconstrained non-Shannon-type inequalities (e.g. \cite{ZY97, ZY98, matus-inf, SixInequalities, projection-method, condineq}).
As an example, we show how to obtain the very first non-Shannon-type inequality using this rule.

\begin{theorem}[Zhang and Yeung, \cite{ZY98}]\label{theo:ZY98}
The following is a $4$-variable information inequality:
\begin{multline*}
I(C\lon D) \le I(C\lon D|A) +  I(C\lon D|B)  +  I(A\lon B) +\\+ I(C\lon D|A) + I(A\lon C|D) + I(A\lon D|C).
\end{multline*}
\end{theorem}
\begin{IEEEproof}
Apply Rule ZY to the following Shannon-type information inequality (which can be verified using a computer program):
\begin{multline*}
I(C\lon D) \le I(C\lon D|A) +  I(C\lon D|B)  +  I(A\lon B) +\\+ I(C\lon D|Z) + I(Z\lon C|D) + I(Z\lon D|C) 
+\\+3I(Z\lon AB|CD).\label{ineq:Shannon5}
\end{multline*}
Let $Z=A$ in the inequality we obtain.
\end{IEEEproof}

\subsection{Makarychev~\emph{et~al} Technique}

\begin{mdframed}[%
        frametitle={Rule MMRV},
        frametitlefont=\scshape\bfseries,
        skipabove=1pt,
        skipbelow= 3pt,
       innerlinewidth=.5pt,
        frametitlerule=true,
        frametitlebackgroundcolor=gray!30, 
        backgroundcolor=gray!10, 
        nobreak=true]
\begin{enumerate}[(A)]
\item
 If we have an information inequality of the form:
 \begin{equation*}
 f(X_\mathcal{N},Y_\mathcal{M}) + g(Y_\mathcal{M},Z) \ge 0;
  \end{equation*}
\item then the following (stronger) inequality is also valid:
 \[
 f(X_\mathcal{N},Y_\mathcal{M}) + g(Y_\mathcal{M},Z) 
 - r_Z H(Z|Y_\mathcal{M}) \ge 0,
 \]
\end{enumerate}
where $r_Z$ is the sum of coefficients of $g$ involving $Z$.
\end{mdframed}
The correctness of this rule is based on a result from the works of Ahlswede\textendash{G\'{a}cs\textendash{K\"{o}rner}}. The general result is presented in a book by Csisz\'{a}r and K\"{o}rner \cite{csiszarkorner}. The relevance of this result was also underlined by Wyner in \cite{Wyner}. We state here a special case suited to our needs.

 \begin{lemma}[Ahlswede\textendash{K\"{o}rner} Lemma, \cite{ahlswedegacskorner,csiszarkorner}]
 \label{lemma:AK}
Let $y_1,\dotsc,y_n,z$ be $n+1$ jointly distributed random variables. Consider their respective $M$ i.i.d. copies $Y_1,\dotsc,Y_n,Z$ . Then there exists a random variable $Z'$ such that:
\begin{itemize}
\item $H(Z'|Y_1,\dotsc,Y_n) = 0$,
\item $H(Y_J|Z') - M\cdot H(y_J|z) =  o(M)$, for all $\varnothing\neq J\subseteq\mathcal{N}$.
\end{itemize}
Denote this $W$ by $AK(Z\lon Y_1,\dotsc,Y_n)$.
\end{lemma} 

\begin{IEEEproof}[Proof of Correctness of \textsc{Rule~MMRV}]
\color{ok}
Consider the joint $M$ i.i.d copies $X^M_\mathcal{N},Y^M_\mathcal{M},Z^M$ of variables $X_\mathcal{N},Y_\mathcal{M},Z$. Let $Z' = AK(Z^M\lon Y^M_\mathcal{M})$ be the variable obtained using Lemma~\ref{lemma:AK}. Apply the inequality of step (A) to the corresponding $M$ independent copies except take $Z=Z'$. Entropy terms not involving $Z'$ are thus $M$ times greater.
Let us compute the entropy terms involving $Z'$ (from $g$) using Lemma~\ref{lemma:AK}:\begin{align*}
H(Z')
&= I(Z'\lon Y^M_\mathcal{M}) + H(Z'|Y^M_\mathcal{M})\\
&= H(Y^M_\mathcal{M}) - H(Y^M_\mathcal{M}|Z') + 0\\
&= M\!\cdot\! H(Y_\mathcal{M}) - M\!\cdot\!  H(Y_\mathcal{M}|Z) + o(M)\\
&= M\!\cdot\! [ H(Z) - H(Z|Y_\mathcal{M})] + o(M).
\end{align*}
Let $J\subseteq\mathcal{M}$, 
\begin{IEEEeqnarray*}{rCl}
H(Z',Y^M_J) 
&=& H(Z') + H(Y^M_J|Z')\\
&=& M\!\cdot\! [ H(Z) - H(Z|Y_{M}) + H(Y_{J}|Z)] + o(M) \\
&=& M\!\cdot\! [H(Z,Y_{J}) - H(Z|Y_{M})] + o(M).
\end{IEEEeqnarray*}
Rewriting our instance of inequality $(A)$ thus gives
\[
M\!\cdot\! [f(X_\mathcal{N},Y_\mathcal{M}) + g(Y_\mathcal{M},Z) - r_ZH(Z|Y_{M})  ]  + o(M) \ge 0,
\]
where $r_Z$ is the sum of coefficients of $g$ involving $Z$.
Dividing the last inequality by $M$ and making $M$ tend to infinity gives the inequality of step $(B)$.
\end{IEEEproof}

As an example, we retrieve Makarychev \emph{et al} proof of the generalization of Zhang and Yeung $4$-variable inequality (see Theorem~\ref{theo:ZY98}).

\begin{theorem}[Makarychev~\emph{et al}, \cite{MMRV}]\label{theo:ineqMMRV}
The following is a $5$-variable information inequality:
\begin{multline*}
I(C\lon D) \le I(C\lon D|A) +  I(C\lon D|B)  +  I(A\lon B) +\\+ I(C\lon D|E) + I(E\lon C|D) + I(E\lon D|C)
\end{multline*}
\end{theorem}
\begin{IEEEproof}
Apply \textsc{Rule~MMRV} to the Shannon-type inequality:
\begin{multline*}
H(Z) \le I(C\lon D|A) +  I(C\lon D|B)  +  I(A\lon B) +\\+ 2H(Z|C) + 2H(Z|D).
\end{multline*}
Let $Z=E$ in the inequality we obtain.
\end{IEEEproof}

Since balancing will appear to be important in the sequel, we state simple properties about the two rules.
 
\begin{proposition}\label{prop:rulebalance}
\leavevmode

\begin{itemize}
\item 
Suppose inequality $(B)$ is inferred from $(A)$ by \textsc{Rule~ZY} and $V$ is a variable,
then
\begin{IEEEeqnarray*}{c}
\mbox{$(A)$ is balanced for $V$ iff $(B)$ is balanced for $V$.}
\end{IEEEeqnarray*}
\item
Suppose inequality $(B)$ is inferred from $(A)$ by \textsc{Rule~MMRV} and $V\neq Z$ is a variable,
then:
\begin{itemize}
\item  $(A)$ is balanced for $V$ iff $(B)$ is balanced for $V$.
\item $(B)$ is balanced for $Z$.
\end{itemize}
\end{itemize}
\end{proposition}
The proof follows immediately from the statements of the rules and the definition of balanced inequalities. Notice that \textsc{Rule~MMRV} is only useful when applied to inequalities that are not balanced for $Z$. However, the rule balances for $Z$ afterwards.

\section{Comparison of Proofs Systems}
\label{sec:comparison}

In the spirit of information inequality provers, we will consider and compare various proof systems based on the two rules described above.

\begin{definition}
A \emph{proof system} (for inequalities) consists of a \emph{pool} $P$ of inequalities and a rule \textsc{T}.
A (computation) \emph{step} in a proof system is described as follows:
\begin{enumerate}
\item Pick an inequality $(A)$ from the convex closure of $P$;
\item Apply rule \textsc{T} to $(A)$ and infer inequality $(B)$;
\item Add $(B)$ to the pool $P$.
\end{enumerate}

A \emph{derivation} is a sequence of valid steps in a system. 
An inequality $(\mathcal{I})$ is \emph{provable in system $S$} if it belongs to the convex closure of the pool of $S$ after a derivation.
\end{definition}

Note that in the previous rules, the naming of the variables is unimportant. The special variable $Z$ may change for each application of a rule.
We want to compare the following systems:

\begin{itemize}
\item \textsc{System ZY}: the system using \textsc{Rule~ZY}.
\item \textsc{System ZY+b}: the system using \textsc{Rule~ZY} and balancing at each step.
\item \textsc{System R}: the system using \textsc{Rule~MMRV}.
\item \textsc{System R+b}: the system using \textsc{Rule~MMRV} and balancing at each step.
\end{itemize}
Usually, a proof system will be initialized with a starting pool of inequalities: the (elemental) Shannon-type inequalities.

First, we show that the two inference rules of Section~\ref{sec:techniques} are in a sense equivalent if we keep in mind Theorem~\ref{theo:balanced} about balanced inequalities.

\begin{theorem}[Equivalence modulo balancing]\label{theo:equivmodbalance}
\leavevmode

Suppose $({B}_1)$ can be inferred from $({A}_1)$ by \textsc{Rule~ZY}, where  $({A}_1)$ is balanced for $Z$.
Then there is an $(A_2)$  such that:
\begin{itemize}
\item $(B_1)$ can be inferred from $({A}_2)$ by \textsc{Rule~MMRV};
\item $(A_2)$ follows from $(A_1)$;
\end{itemize}

Suppose $({B'}_1)$ can be inferred from $({A'}_1)$ by \textsc{Rule~MMRV}. 
Then there is an $(A'_2)$ such that: 
\begin{itemize}
\item $({B'}_1)$ can be inferred from $({A'}_2)$ by \textsc{Rule~ZY};  
\item $(A'_2)$  balances for $Z$ into $(A'_1)$.
\end{itemize}
\end{theorem}
\begin{IEEEproof}
\leavevmode

\noindent
\textsc{Rule~ZY} $\Ra$ \textsc{Rule~MMRV}: 
Suppose
 \begin{equation}
 f(X_\mathcal{N},Y_\mathcal{M}) + g(Y_\mathcal{M},Z) + 
 \alpha I(Z\lon X_\mathcal{N}|Y_\mathcal{M})\ge 0,
 \label{ineq:A1}\tag{$A_1$}
   \end{equation}
 for some $\alpha\ge0$, is a valid information inequality. 
 By \textsc{Rule~ZY}, the stronger
 \begin{equation}
  f(X_\mathcal{N},Y_\mathcal{M}) + g(Y_\mathcal{M},Z) \ge0
  \label{ineq:B1}\tag{$B_1$}
  \end{equation}
  is also valid. 
Let us show that inequality~\eqref{ineq:B1} can also be obtained using \textsc{Rule~MMRV}. and balancing. Start from the inequality
 \begin{equation}
 f(X_\mathcal{N},Y_\mathcal{M}) + g'(Y_\mathcal{M},Z) \ge 0
 \label{ineq:A2}\tag{$A_2$}
   \end{equation}
   defined using $g' = g + \alpha H(Z|Y_\mathcal{M})$. This inequality is valid since $\alpha$ is non-negative, thus \eqref{ineq:A2} follows from \eqref{ineq:A1} and $H(Z|Y_\mathcal{M}) \ge I(Z\lon X_\mathcal{N}|Y_\mathcal{M})$.
 
By applying \textsc{Rule~MMRV} we get 
\begin{equation}
 f(X_\mathcal{N},Y_\mathcal{M}) + g'(Y_\mathcal{M},Z) - r'_Z H(Z|Y_\mathcal{M})\ge0,
 \label{ineq:B2}\tag{$B_2$}
  \end{equation}
where $r'_Z$ is the sum of coefficients of $g'$ involving $Z$. By definition of $g'$ we have $r'_Z = \alpha + r_Z$, where $r_Z$ is the sum of coefficients of $g$ involving $Z$.
Thus inequality~\eqref{ineq:B2} rewrites to
\[
 f(X_\mathcal{N},Y_\mathcal{M}) + g(Y_\mathcal{M},Z) - r_Z H(Z|Y_\mathcal{M}) \ge0,
\]
and since \eqref{ineq:A1} is balanced for $Z$, \ie $r_Z = 0$, the previous inequality is exactly inequality~\eqref{ineq:B1}.
\medskip 

\noindent
\textsc{Rule~MMRV} $\Ra$ \textsc{Rule~ZY}:
Suppose
\begin{equation}
 f(X_\mathcal{N},Y_\mathcal{M}) + g(Y_\mathcal{M},Z)  \ge 0
 \label{ineq:A2p}\tag{$A'_2$}
\end{equation}
 is a valid information inequality. By \textsc{Rule~MMRV}, the stronger
\begin{equation}
  f(X_\mathcal{N},Y_\mathcal{M}) + g(Y_\mathcal{M},Z) - r_Z H(Z|Y_\mathcal{M}) \ge0
\label{ineq:B2p}\tag{$B'_2$}
\end{equation}
  is also valid. Let us show that inequality~\eqref{ineq:B2p} can also be inferred using \textsc{Rule~ZY} and balancing.
  
Notice first that \[H(Z|Y_\mathcal{M}) = H(Z|X_\mathcal{N}Y_\mathcal{M}) + I(Z\lon X_\mathcal{N}|Y_\mathcal{M}),\] therefore
\eqref{ineq:A2p} rewrites to 
\begin{multline*}
 f(X_\mathcal{N},Y_\mathcal{M}) + [g(Y_\mathcal{M},Z) - r_Z H(Z|Y_\mathcal{M})] +\\+ r_Z  H(Z|X_\mathcal{N}Y_\mathcal{M}) + r_Z I(Z\lon X_\mathcal{N}|Y_\mathcal{M}) \ge 0,
 \end{multline*}
 where $r_Z$ is the sum of the coefficients of $g$ involving $Z$.
Balancing this inequality for $Z$ gives:
\begin{multline}
 f(X_\mathcal{N},Y_\mathcal{M}) + [g(Y_\mathcal{M},Z)-r_Z H(Z|Y_\mathcal{M})] +\\+ r_Z I(Z\lon X_\mathcal{N}|Y_\mathcal{M})  \ge 0
 \label{ineq:A1p}\tag{$A'_1$}
\end{multline}
Applying the inference rule of \textsc{Rule~ZY} to \eqref{ineq:A1p} gives
\begin{equation}
 f(X_\mathcal{N},Y_\mathcal{M}) + g(Y_\mathcal{M},Z) - r_Z H(Z|Y_\mathcal{M}) \ge0,
 \label{ineq:B1p}\tag{$B'_1$}
\end{equation}
which is exactly inequality~\eqref{ineq:B2p}.
\end{IEEEproof}

This result shows the importance of balancing non-Shannon-type inequalities.
For a Shannon-type inequality, its balanced counterpart is also Shannon-type and thus already belongs to the pool. However, the balanced counterpart of a non-Shannon-type inequality may not belong to the pool.

\begin{corollary}
Let $(\mathcal{I})$ be an information inequality.
The following are equivalent:
\begin{itemize}
\item $(\mathcal{I})$ is provable in  \textsc{System R+b}.
\item $(\mathcal{I})$ is provable in  \textsc{System ZY+b}.
\item $(\mathcal{I})$ is provable in  \textsc{System R} when using only inequalities balanced for all variables but Z.
\item $(\mathcal{I})$ is provable in  \textsc{System ZY} when using only balanced for $Z$ inequalities.
\end{itemize}
\end{corollary}
\begin{IEEEproof}
Follows immediately from Proposition~\ref{prop:rulebalance} and Theorem~\ref{theo:equivmodbalance}.
\end{IEEEproof}

\section{Conclusion}

We have shown that it does not matter which of the two rules to implement in an information inequality prover, as long as it applies them to balanced inequalities. Since the cost of checking and balancing an inequality is minor, and balanced inequalities are stronger than their counterparts, they should be useful for such programs. Moreover, we have seen that balancing is not compulsory at each step because the rules can only improve balancing (Proposition~\ref{prop:rulebalance}). A last argument in favour of balanced inequalities might be the fact that they are the only inequalities valid for continuous entropy (see \cite[Theorem~2]{chan-balanced}).



\bibliography{ref} 

\end{document}